# Stages of destruction and elastic compression of granular nanoporous carbon medium at high pressures

I.M. Neklyudov, O.P. Ledenyov, N.B. Bobrova, A.A. Chupikov

*Institute of Solid State Physics, Materials Science and Technologies, National Scientific Centre Kharkov Institute of Physics and Technology, Academicheskaya 1, Kharkov 61108, Ukraine.*

The granular nanoporous carbon medium, made of the cylindrical coal granules of the adsorbent of CKT-3, at an influence by the high pressures from 1 MPa to 3 GPa has been researched. The eight consecutive stages of the material's specific volume change, which is characterized by a certain dependence of the volume change on the pressure change, have been registered. It is shown that there is a linear dependence on the double log-log plot of the material's specific volume change on the pressure for an every stage of considered process. The two stages are clearly distinguished: a stage of material's mechanical destruction, and a stage of elastic compression of material without the disintegration of structure at a nano-scale. The hysteresis dependence of the material's specific volume change on the pressure change at the pressure decrease is observed. The small disperse coal dust particles jettisoning between the high pressure cell and the base plate was observed, resulting in the elastic stress reduction in relation to the small disperse coal dust particles volume. The obtained research data can be used to improve the designs of air filters for the radioactive chemical elements absorption at the NPP with the aims to protect the environment.

## Introduction

The functional materials with the special physical properties in a certain range of technical parameters such as the temperatures, pressures, deformations are used in the modern engineering to design a number of various devices. However, the selected physical properties of functional materials at the operational limits are not well researched.

The granular nanoporous carbon medium, made of the cylindrical coal granules of the adsorbent of CKT-3, at an influence by the high pressures from 1 MPa to 3 GPa represents a main research subject in this paper. The granular nanoporous carbon medium, made of the cylindrical coal granules of the adsorbent of CKT-3 is used in the air filters for the radioactive chemical elements and their isotopes absorption at the nuclear power plants (NPP) with the aims to defend the working personal from the harsh operational environment.

In [1], the research results on the distribution of the radioactive chemical elements and their isotopes along the air filters after the long term operation at the NPP with the application of the gamma resonant spectroscopy method has been completed. The air filters have been used in the operation for a number of years, and the accumulated chemical elements and their isotopes had a non-uniform distribution with a number of maxima of concentration density.

It was necessary to create a new theory, considering an important role by the forced acoustic resonant oscillations with the purpose to explain an appearance of the concentration density maxima in the granular nanoporous carbon medium, made of the cylindrical coal granules of the adsorbent of CKT-3, in the air filter [2].

It was observed that the primary accumulation of the absorbed chemical elements and their isotopes occurs in the antinodes positions of the forced acoustic oscillations, having a distribution in the form of a series of concentration density maxima along the length of the granular nanoporous carbon medium, made of the cylindrical coal granules of the adsorbent of CKT-3, in the air filter [2].

The small disperse coal dust particles appear as a result of an abrasion of mechanical contacts between the coal granules at the air pressure fluctuations during the air flow in an air filter, and precipitate at the same positions along the length of the granular nanoporous carbon medium in the core of an air filter. The small disperse coal dust particles can intensify the process of the radionuclides absorption, because the path length by the radionuclides to the pores of small dimensions in the small disperse coal dust particles is relatively short. Therefore, a physical chemical role by the small disperse coal dust particles fraction in the process of absorption of the chemical elements has to be considered comprehensively. In addition, it makes sense to note that the small disperse coal dust particles fraction plays an essential role in the space physics [3], plasma physics [4], environment science, etc.

Let us emphasis that the mechanical strength of cylindrical coal granules is precisely characterized by the disintegration property due to an abrasion process mainly [5]. Therefore, there was a certain interest to conduct a research on the mechanical durability and dense packing of the granular nanoporous carbon medium with the cylindrical coal granules with the fractal structure of pores at an application of a wide range of high pressures [6].

In this research, we would like to clarify the following problems: 1) destruction of the macro-pores in the granular carbon medium with nanoporous and 2) elastic compression of the nano-pores in the small disperse coal dust particles in the carbon medium as a result of an application of high pressures.

## Samples characteristics and measurements technique

The granular nanoporous carbon medium, made of the cylindrical coal granules of the adsorbent of CKT-3, at an application by the high pressures from 1 MPa up to 3 GPa has been researched. The coal granules had the cylindrical geometrical form with the diameter of ~ 2 mm and the length of ~ 3.2 mm. The measured material density is $\rho_{gr} \approx 0.9213$ g/cm$^3$, that is approximately ~ 0.4 of the graphite density, $\rho_C = 2,253$ g/cm$^3$.

The bulk filling of air filter by the cylindrical coal granules can generally be from ~ 62 % up to ~ 75 %, and it depends on both 1) the geometrical shapes of granules and impurities as well as 2) a relation between the empty space volume and the characteristic sizes of granules.

In our experiment, the cylindrical coal granules are filed in the cylindrical volume. The diameter of cylindrical coal granule is in 4 times smaller than the diameter of cylindrical volume. The length of cylindrical coal granule is in 2.5 times smaller than the length of cylindrical volume. It results in a situation, when the filing density of the cylindrical volume by the cylindrical coal granules is decreased. In the researched case, the density of bulk filed cylindrical coal granules was equal to $\rho_{bulk} = 0.567$ g/cm$^3$, that is 62% of volumetric filing.

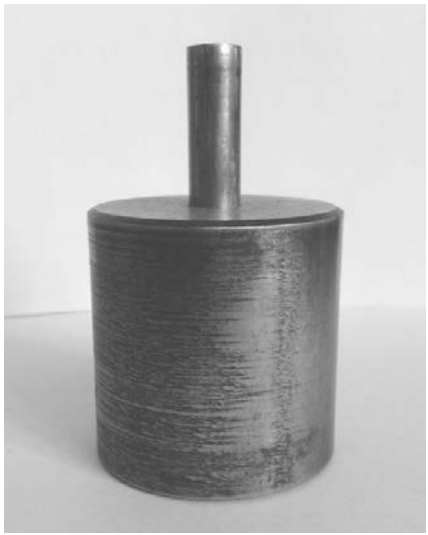

*Fig. 1. High pressure cell.*

It is known that cylindrical coal granules of CKT-3 have the macro-pores with the diameter of ~ 100-200 nm, the volume of macro-pores is 0.2-0.8 cm$^3$/g, the specific surface of macro-pores is 0.5-2 m$^2$/g [7]. The transitional nano-pores have the sizes from 1.6 nm to 100 nm and the specific area of 70 m$^2$/g. The smallest nano-pores have the sizes from 0.6–0.7 up to 1.6 nm, and the specific volume of 0,2–0,6 cm$^3$/g, and the big effective area from 700 up to 1500 m$^2$/g.

Thus, it is necessary to highlight a fact that the absorption process takes place in the nano-pores mainly, where the adsorbed atoms of radioactive chemical elements and their isotopes are captured in the volumes of nano-pores. The macro-pores in the cylindrical coal granules in the granular nanoporous carbon medium play a role of transporting channels.

The high pressure cell represents a cylindrical chamber with the diameter of 40 mm and the height of 40 mm with an internal hole with the diameter of 8 mm and a punch with the diameter of 8 mm and the length of 40 mm made of the high-strength tempered steel as depicted in Fig. 1.

The external pressure was created by the hydraulic press, which can generate the high pressures. The multiplication coefficient in a high pressure cell is 100, allowing to create the high pressures up to 3 GPa.

## Discussion on experimental results

The change of volume of the granular nanoporous carbon medium, made of the cylindrical coal granules of the adsorbent of CKT-3, at an influence by the high pressures from 1 MPa to 3 GPa is measured.

The experimental results on the change of volume of the granular nanoporous carbon medium, made of the cylindrical coal granules of the adsorbent of CKT-3, at an influence by the high pressures from 1 MPa to 3 GPa are presented in Fig. 2.

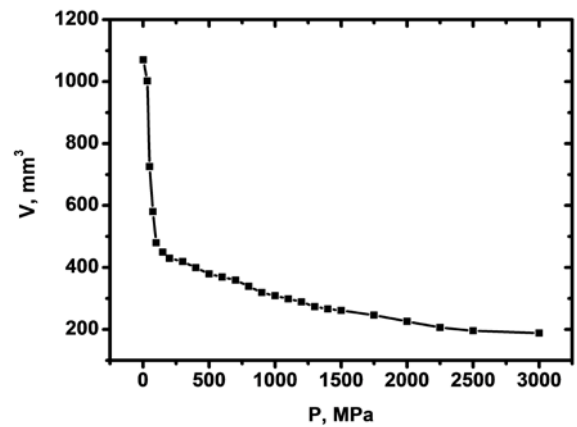

*Fig. 2. Dependence of volume on pressure for granular nanoporous carbon medium.*

It can be seen that the volume, occupied by the cylindrical coal granules, quickly enough decreases at an increase of the pressure up to ~100 MPa, and then the volume change is slowing down.

At the maximum pressure of 3000 MPa, the volume of the granular nanoporous carbon medium decreases in more than 5 times, comparing to the initial volume.

For more detailed research of dependence V (P), it has been constructed on the log-log scale (Fig. 3).

In this case, it can be seen more accurately that the volume of the granular nanoporous carbon medium changes at an influence by the applied pressure. The obtained dependence can be divided into the eight stages in which the every dependence V (P) can be presented as a close to the linear dependence on the log-log scale.

In the range of small applied pressures from 1 MPa to 32 MPa (the area 1 in Fig. 3), there is the volume change of ~ 6.6 %, which can be connected with the packing of a bulk layer of cylindrical coal granules without their mechanical destruction.

The following stage of process (2) is probably connected with the destruction of a macrostructure of the cylindrical coal granules, and the volume of sample at 100 MPa changes on 49 % from the initial volume at the pressure of 1 MPa.

In the subsequent stages (3-7), the deformation process influences the volumes of pores inside the cylindrical coal granules in the granular nano-porous carbon medium.

The volume change at all these stages is almost twice less, than at the stage (2), and it is 25.65 % of the initial volume of researched sample.

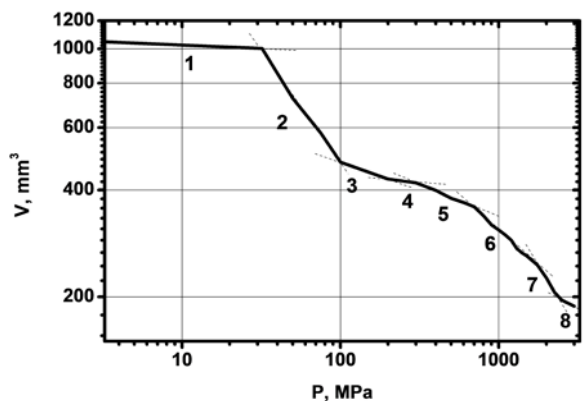

*Fig.3. Log-log experimental dependence for bulk sample of volume V as function of pressure P.*

At the stage (8), the deformation of sample at the high pressures up to 3 GPa has place. This process is accompanied by a small change of volume of sample on ~ 3.8 %.

For more evident estimation of changes of volume of sample at each stage of the deformation process, the dependence of the relative change of volume in the percentage of the previous stage, related to the change of pressure on 1 MPa, has been created in Fig 4.

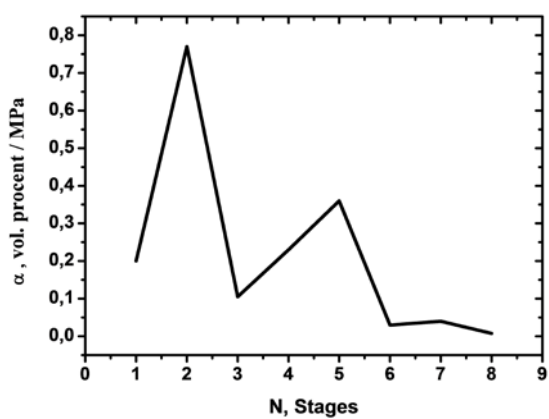

*Fig. 4. Coefficient of change of volume in percentage at change of pressure on 1 MPa related to each stage.*

As it is visible in Fig. 4, the biggest compression of sample is related to a change of pressure on 1 MPa and it occurs at the stages 2 and 5. The small relative maximum is observed at the 7$^{th}$ stage. It is possible to believe that the destruction of the cylindrical coal granules takes place at these stages of the deformation process at the relatively small pressures (stage 2). The stage 5 is connected with the destruction of macro-pores in the adsorbent, and the stage 7 is responsible for a partial destruction of the bigger micro-pores.

We believe that the nano-pores in the samples do not disintegrate, but they are only elastically compressed. During this process, the high energy of elastic deformation can be accumulated in the sample at the applied high pressures.

Let's highlight the fact that the cylindrical coal granules are fractured and destructed in the researched granular nanoporous carbon medium at an increase of external applied pressure. This process is accompanied by both 1) an increase of applied pressure on the small disperse coal dust particles and 2) an increase of the elastic compression of material of which they consist.

It is known [8] that the processes of the deformation and the distribution of forces in the granulated soft condensed matter occur differently, than in the elastic condensed matter. Therefore, at a decrease of external pressure, the small disperse nano-particles are in a jammed state, and the pressure release in the sample does not occur, as it would be in the case of an elastic condensed matter sample.

Indeed, as it was observed during the experiment, at pressure decrease from 3 GPa down to ~ 100 MPa, the volume of the compressed sample increases a little, but at the specified pressures, the dust of small disperse coal nano-particles appeared between the cell and the metal plate on which it is based (Fig. 5). It shows that the small disperse coal dust particles medium in a high pressure cell was elastically compressed, and at pressure release, it appeared between the the cell and the metal plate on which it is fixed.

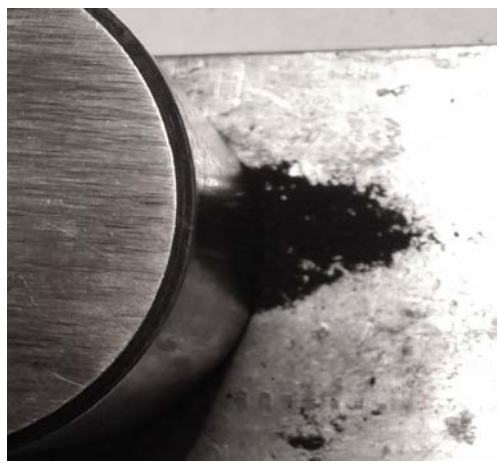

*Fig. 5. Small disperse dust jettisoning from high pressure cell.*

After this, the pressure was completely decreased, but the small disperse coal dust in the cell's channel was in a packed state, and it was necessary to apply a certain additional pressure take it out.

## Conclusion

The granular nanoporous carbon medium, made of the cylindrical coal granules of the adsorbent of CKT-3, at an influence by the high pressures from 1 MPa to 3 GPa has been researched. The eight consecutive stages of the material's specific volume change, which is characterized by a certain dependence of the volume change on the pressure change, have been registered. It is shown that there is a linear dependence lgV (lgP) on the double log-log plot of the material's specific volume change on the pressure for an every stage of considered process. The two stages of mechanical transformation of researched sample are clearly distinguished: 1) a stage of material's mechanical destruction, and 2) a stage of elastic compression of material without the subsequent disintegration of internal structure at a nano-scale. The magnitudes of external applied pressures of fastest deformation process of cylindrical coal granules are measured. The presence of elastic deformations of the granular nanoporous carbon medium at which the nano-pores in the granular nanoporous carbon medium do not destruct is found. The small disperse coal dust particles jettisoning between the high pressure cell and the base plate was observed, resulting in the elastic stress reduction in relation to the small disperse coal dust particles volume. The obtained research data can be used to improve the designs of air filters for the chemical elements absorption at the NPP.

This innovative research is completed in the frames of the nuclear science and technology fundamental research program, facilitating the environment protection from the radioactive contamination at the nuclear power plants with the fast neutrons nuclear reactors and the thermonuclear reactors at the *National Scientific Centre Kharkov Institute of Physics and Technology (NSC KIPT)* in Kharkov in Ukraine.

The research is funded by the *National Academy of Sciences in Ukraine (NASU)*.

This research article was published in the *Problems of Atomic Science and Technology* (*VANT*) in [9] in 2015.

*
 E-mail: ledenyov@kipt.kharkov.ua